\documentclass[proceedings]{JHEP3}
\usepackage{amsfonts}
\usepackage{amssymb}
\usepackage{amsfonts}
\usepackage{amsmath}
\usepackage{epsfig}

\newcommand{\be}{\begin{equation}}
\newcommand{\en}{\end{equation}}
\newcommand{\bea}{\begin{eqnarray}}
\newcommand{\ena}{\end{eqnarray}}
\newcommand{\beano}{\begin{eqnarray*}}
\newcommand{\enano}{\end{eqnarray*}}
\newcommand{\bee}{\begin{enumerate}}
\newcommand{\ene}{\end{enumerate}}

\newcommand{\Lc}{{\cal L}}

\conference{Non self-adjoint model with unbound metric}
\abstract{We demonstrate that a non self-adjoint Hamiltonian of  harmonic  oscillator type defined on a two-dimensional noncommutative space can be
diagonalized exactly by making use of pseudo-bosonic operators. The model
admits an antilinear symmetry and is of the type studied in the context of
PT-symmetric quantum mechanics. Its eigenvalues are computed to be real for the entire range of the coupling constants and the biorthogonal sets of eigenstates for the Hamiltonian and its adjoint are explicitly constructed. We show that despite the fact that these sets are complete and biorthogonal, they involve an unbounded metric operator and therefore do not constitute (Riesz) bases for the Hilbert space $\Lc^2(\Bbb R^2)$, but instead only D-quasi bases. As recently proved by one of us (FB), this is sufficient to deduce several interesting consequences.}

\title{A non self-adjoint model on a two dimensional noncommutative space
with unbound metric}
\author{Fabio Bagarello$^{\circ}$ and Andreas Fring$^\bullet$ \\
$^\circ$ Dipartimento di Energia, Ingegneria dell'Informazione e Modelli
Matematici, \\
$\,\,$ Facolt\`{a} di Ingegneria, Universit\`{a} di Palermo, I-90128
Palermo, Italy and\\
$\,\,$ INFN, Sezione di Torino, Italy\\
$^\bullet$ Department of Mathematical Science, City University London,\\
$\,\,$ Northampton Square, London EC1V 0HB, UK\\
E-mail: fabio.bagarello@unipa.it, a.fring@city.ac.uk}

\input{tcilatex}
\begin{document}

\section{Introduction}

In the last 15 years more and more of physicists and mathematicians have
developed an interest in non-Hermitian and non self-adjoint operators
possessing real eigenvalues. Such type of models have been investigated
before, but the more recent interest has been initiated by the seminal paper 
\cite{Bender:1998ke} in which the complex cubic potential and its close
relatives have been studied. The original considerations, focussing mainly
on the aspect of the possibility to formulate consistent quantum mechanical
systems, have broadened quickly and are partly replaced by a more general
analysis of related aspects. Many experiments \cite{Ex1,Ex2,Ex3,Ex4} have
now been carried out, mainly for optical analogues to the quantum mechanical
systems, exploiting $\mathcal{PT}$-symmetric phase transitions where real
eigenvalues merge into two complex conjugate pairs, to obtain gain and loss
structures. We refer the reader to \cite{ben,mosta,znorev} for some reviews
on what is commonly named \emph{quasi-Hermitian} \cite{Dieu,Urubu}, \emph{%
pseudo-Hermitian} \cite{pseudo1,Mostafazadeh:2002hb} or $\mathcal{PT}$-\emph{%
symmetric} \cite{EW,Bender:1998ke} quantum mechanics. However, it was
recently pointed out by Krejcirik and Siegl \cite{petr} that more
mathematically oriented treatments of these type of Hamiltonians are
required, as for instance the complex cubic potential lacks to posses a
Riesz basis of eigenstates. Therefore we can still not associate a standard
quantum mechanical interpretation to this model. The purpose of this paper
is to shed more light on these issues.

Modifying recent ideas \cite{Tri}, one of us has newly introduced the notion
of ${\mathcal{D}}$-pseudo bosons (${\mathcal{D}}$-pbs), \cite{bagnewpb}, and
used them in connection with several physical systems, whose Hamiltonians
are non self-adjoint operators, \cite{margh}. Among other aspects, it was
shown that ${\mathcal{D}}$-pbs could be useful in the analysis of a
two-dimensional harmonic oscillator described by the Hamiltonian 
\begin{equation}
\hat{H}=\frac{1}{2}(\hat{p}_{1}^{2}+\hat{x}_{1}^{2})+\frac{1}{2}(\hat{p}%
_{2}^{2}+\hat{x}_{2}^{2})+i\left[ A(\hat{x}_{1}+\hat{x}_{2})+B(\hat{p}_{1}+%
\hat{p}_{2})\right] ,  \label{1}
\end{equation}%
where $(\hat{x}_{j},\hat{p}_{j})$ are noncommutative operators satisfying $[%
\hat{x}_{j},\hat{p}_{k}]=i\delta _{j,k}1\!\!1$, $[\hat{x}_{j},\hat{x}%
_{k}]=i\theta \epsilon _{j,k}1\!\!1$, $[\hat{p}_{j},\hat{p}_{k}]=i\tilde{%
\theta}\epsilon _{j,k}1\!\!1$, where $\theta $ and $\tilde{\theta}$ are two
real small parameters, measuring the non commutativity of the system. In 
\cite{margh} a perturbative expansion in $\theta $ and $\tilde{\theta}$ was
set up and it was shown, in particular, that if one neglects all the terms
which are at least quadratic in $\theta $ and $\tilde{\theta}$ we can
construct explicitly the eigenvectors of (the approximated version of) $\hat{%
H}$ and deduce the related eigenvalues.

In this paper we show that, if the non commutativity is restricted to the
spatial variables only, i.e. if $\tilde{\theta}=0$, then $\hat{H}\,$, and a
slightly generalized version of it, can be exactly diagonalized in terms of $%
{\mathcal{D}}$-pbs. The corresponding eigenbases are biorthonormal, but
involve a metric operator that is unbounded, together with its inverse. Thus
we will draw a similar conclusion as reached in \cite{petr} and, more
recently, in \cite{bag2013august}.

It may be worth to underline that these results, all together, suggest that
several common believes usually taken for granted in the physical literature
on these topics require some more care than usually adopted. For instance,
in \cite{probl1} (as well as in many other papers, \cite{fbbook}), the
biorthogonal sets of eigenstates of a rather general $H$, with $H^{\dagger
}\neq H$, are used to produce a resolution of the identity. In other words,
they are used as bases in the Hilbert space. However, the results in \cite%
{petr,bag2013august}, and those given in this paper, show that this is not
always possible, even for extremely simple models. This, we believe, helps
clarifying the situation, showing that many claims need to be analyzed in
more details.

This article is organized as follows: in the next section we review the
definition and a few central results on ${\mathcal{D}}$-pbs. In section 3 we
introduce the 2d-harmonic oscillator with linear term in the momenta and
position on a noncommutative flat space and we analyze it in terms of ${%
\mathcal{D}}$-pbs. We provide the computation of how it may be written in
terms of ${\mathcal{D}}$-pb number operators and subsequently we verify the
underlying assumptions, needed to have something more than just a formal
theory. This will allow for the construction of biorthonormal sets, which
are, however, shown not to be Riesz bases and not even bases, but just ${%
\mathcal{D}}$-quasi bases. Our conclusions are stated in section 4.

\section{Pseudo-bosons, generalities}

We briefly review here few definitions and central properties of ${\mathcal{D%
}}$-pbs. More details can be found in \cite{bagnewpb}.

Let $\mathcal{H}$ be a given Hilbert space with scalar product $\left\langle
.,.\right\rangle $ and related norm $\Vert .\Vert $. Furthermore, let $a$
and $b$ be two operators acting on $\mathcal{H}$, with domains $D(a)$ and $%
D(b)$ respectively, $a^{\dagger }$ and $b^{\dagger }$ their respective
adjoints, and let ${\mathcal{D}}$ be a dense subspace of $\mathcal{H}$ such
that $a^{\sharp }{\mathcal{D}}\subseteq {\mathcal{D}}$ and $b^{\sharp }{%
\mathcal{D}}\subseteq {\mathcal{D}}$, where $x^{\sharp }$ is $x$ or $%
x^{\dagger }$. It is worth noticing that we are not requiring here that ${%
\mathcal{D}}$ coincides with either $D(a)$ or $D(b)$. Nevertheless, for
obvious reasons, ${\mathcal{D}}\subseteq D(a^{\sharp })$ and ${\mathcal{D}}%
\subseteq D(b^{\sharp })$.

\noindent \textbf{Definition:} The operators $(a,b)$ are ${\mathcal{D}}$%
-pseudo-bosonic if, for all $f\in {\mathcal{D}}$, we have 
\begin{equation}
a\,b\,f-b\,a\,f=f.  \label{31}
\end{equation}

Sometimes, to simplify the notation, instead of (\ref{31}) we will simply
write $[a,b]=1 \!\! 1$, having in mind that both sides of this equation have
to act on $f\in{\mathcal{D}}$.

\vspace{2mm}

Our working assumptions are the following:

\vspace{2mm}

\noindent \textbf{Assumption ${\mathcal{D}}$-pb 1: }There exists a non-zero $%
\varphi _{0}\in {\mathcal{D}}$ such that $a\,\varphi _{0}=0$.

\vspace{1mm}

\noindent \textbf{Assumption ${\mathcal{D}}$-pb 2: }There exists a non-zero $%
\Psi _{0}\in {\mathcal{D}}$ such that $b^{\dagger }\,\Psi _{0}=0$.

\vspace{2mm}

Then, if $(a,b)$ satisfy the above definition, it is obvious that $\varphi
_{0}\in D^{\infty }(b)$ and that $\Psi _{0}\in D^{\infty }(a^{\dagger })$,
with $D^{\infty }(x)$ denoting the common domain of all powers of $x$. Thus
we can define the following vectors, all belonging to ${\mathcal{D}}$: 
\begin{equation}
\varphi _{n}:=\frac{1}{\sqrt{n!}}\,b^{n}\varphi _{0},\qquad \Psi _{n}:=\frac{%
1}{\sqrt{n!}}\,{a^{\dagger }}^{n}\Psi _{0},  \label{32}
\end{equation}%
for $n\geq 0$. As in \cite{bagnewpb} we introduce the sets $\mathcal{F}%
_{\Psi }=\{\Psi _{n},\,n\geq 0\}$ and $\mathcal{F}_{\varphi }=\{\varphi
_{n},\,n\geq 0\}$. Once again, since ${\mathcal{D}}$ is stable under the
action of $a^{\sharp }$ and $b^{\sharp }$, we deduce that each $\varphi _{n}$
and each $\Psi _{n}$ belongs to the domains of $a^{\sharp }$, $b^{\sharp }$
and $N^{\sharp }$, where $N:=ba$.

It is now straightforward to deduce the following lowering and raising
relations:%
\begin{equation}
\begin{array}{llll}
~a\varphi _{n}=\sqrt{n}\,\varphi _{n-1},~a\,\varphi _{0}=0,~~~ &  & 
b^{\dagger }\Psi _{n}=\sqrt{n}\,\Psi _{n-1},~b^{\dagger }\Psi _{0}=0,~~ & 
\text{for }n\geq 1,\text{ } \\ 
a^{\dagger }\Psi _{n}=\sqrt{n+1}\Psi _{n+1}, &  & ~b\,\varphi _{n}=\sqrt{n+1}%
\varphi _{n+1}, & \text{for }n\geq 0,\text{ }%
\end{array}
\label{33}
\end{equation}%
as well as the following eigenvalue equations: $N\varphi _{n}=n\varphi _{n}$
and $N^{\dagger }\Psi _{n}=n\Psi _{n}$ for $n\geq 0$. As a consequence of
these equations, choosing the normalization of $\varphi _{0}$ and $\Psi _{0}$
in such a way $\left\langle \varphi _{0},\Psi _{0}\right\rangle =1$, we
deduce that 
\begin{equation}
\left\langle \varphi _{n},\Psi _{m}\right\rangle =\delta _{n,m},  \label{34}
\end{equation}%
for all $n,m\geq 0$. The third assumption originally introduced in \cite%
{bagnewpb} is the following:

\vspace{2mm}

\noindent \textbf{Assumption ${\mathcal{D}}$-pb 3: } $\mathcal{F}_{\varphi }$
is a basis for $\mathcal{H}$.

\vspace{1mm}

This is equivalent to the request that $\mathcal{F}_{\Psi }$ is a basis for $%
\mathcal{H}$ as well, \cite{bagnewpb}. In particular, if $\mathcal{F}%
_{\varphi }$ and $\mathcal{F}_{\Psi }$ are Riesz bases for $\mathcal{H}$,
the ${\mathcal{D}}$-pbs were called \emph{regular}.

In \cite{bagnewpb} also a weaker version of Assumption ${\mathcal{D}}$-pb 3
has been introduced, useful for concrete physical applications: for that,
let $\mathcal{G}$ be a suitable dense subspace of $\mathcal{H}$. Two
biorthogonal sets $\mathcal{F}_{\eta }=\{\eta _{n}\in \mathcal{G},\,g\geq 0\}
$ and $\mathcal{F}_{\Phi }=\{\Phi _{n}\in \mathcal{G},\,g\geq 0\}$ were
called \emph{$\mathcal{G}$-quasi bases} if, for all $f,g\in \mathcal{G}$,
the following holds: 
\begin{equation}
\left\langle f,g\right\rangle =\sum_{n\geq 0}\left\langle f,\eta
_{n}\right\rangle \left\langle \Phi _{n},g\right\rangle =\sum_{n\geq
0}\left\langle f,\Phi _{n}\right\rangle \left\langle \eta
_{n},g\right\rangle.  \label{35}
\end{equation}%
Is is clear that, while Assumption ${\mathcal{D}}$-pb 3 implies (\ref{35}),
the reverse is false. However, if $\mathcal{F}_{\eta }$ and $\mathcal{F}%
_{\Phi }$ satisfy (\ref{35}), we still have some (weak) form of resolution
of the identity. Now Assumption ${\mathcal{D}}$-pb 3 is replaced by the
following:

\vspace{2mm}

\noindent \textbf{Assumption ${\mathcal{D}}$-pbw 3: }$\mathcal{F}_{\varphi }$
and $\mathcal{F}_{\Psi }$ are $\mathcal{G}$-quasi bases.

\vspace{1mm}

Let now assume that Assumption ${\mathcal{D}}$-pb 1, ${\mathcal{D}}$-pb 2,
and ${\mathcal{D}}$-pbw 3 are satisfied, with $\mathcal{G}={\mathcal{D}}$,
and let us consider a self-adjoint, invertible, operator $\Theta$, which
leaves, together with $\Theta^{-1}$, ${\mathcal{D}}$ invariant: $\Theta{%
\mathcal{D}}\subseteq{\mathcal{D}}$, $\Theta^{-1}{\mathcal{D}}\subseteq{%
\mathcal{D}}$. Then, as in \cite{bagnewpb}, we say that $(a,b^\dagger)$ are $%
\Theta-$conjugate if $af=\Theta^{-1}b^\dagger\,\Theta\,f$, for all $f\in{%
\mathcal{D}}$. Moreover, we can check that, for instance, $(a,b^\dagger)$
are $\Theta-$conjugate if and only if $(b,a^\dagger)$ are $\Theta-$conjugate
and that, assuming that $\left<\varphi_0,\Theta\varphi_0\right>=1$, $%
(a,b^\dagger)$ are $\Theta-$conjugate if and only if $\Psi_n=\Theta\varphi_n$%
, for all $n\geq0$. Finally, if $(a,b^\dagger)$ are $\Theta-$conjugate, then 
$\left<f,\Theta f\right>>0$ for all non zero $f\in {\mathcal{D}}$. The
details of these proofs can be found in \cite{bag2013august}. Notice also
that, not surprisingly, we also deduce that $Nf=\Theta^{-1}N^\dagger\Theta f$%
, for all $f\in {\mathcal{D}}$.

\section{Noncommutative two dimensional harmonic oscillator with linear terms%
}

Let us now consider the non self-adjoint two dimensional harmonic oscillator
with linear terms in the momenta and positions 
\begin{equation}
\tilde{H}=\frac{1}{2m}(\tilde{p}_{1}^{2}+\tilde{p}_{2}^{2})+\frac{m\omega
^{2}}{2}(\tilde{x}_{1}^{2}+\tilde{x}_{2}^{2})+i\alpha _{1}\tilde{x}%
_{1}+\alpha _{2}\tilde{x}_{2}+\alpha _{3}\tilde{p}_{1}+i\alpha _{4}\tilde{p}%
_{2},  \label{HL}
\end{equation}%
on the noncommutative flat space with the nonvanishing commutators $[\tilde
x_{1},\tilde x_{2}]=i\theta $, $[\tilde x_{j},\tilde p_{j}]=i\hbar $ for $%
j=1,2.$ Here $\theta $ and $\alpha _{i}$ for $i=1,2,3,4$ are real
dimensionful parameters. Note that this Hamiltonian is non self-adjoint even
when viewed on a standard space. However, $\tilde{H}$ is constructed in such
a way that it is left invariant with respect to the antilinear symmetry $%
\mathcal{PT}_{-}$: $\tilde{x}_{1}\rightarrow -\tilde{x}_{1}$, $\tilde{x}%
_{2}\rightarrow \tilde{x}_{2}$, $\tilde{p}_{1}\rightarrow \tilde{p}_{1}$, $%
\tilde{p}_{2}\rightarrow -\tilde{p}_{2}$ and $i\rightarrow -i$ \cite{DFG}.
Thus in the general spirit of $\mathcal{PT}$-symmetric quantum mechanics 
\cite{EW,Bender:1998ke} the Hamiltonian is guaranteed to have real
eigenvalues provided that its eigenfunctions are eigenstates of $\mathcal{PT}%
_{-}$. Evidently in atomic units, $m=\omega =\hbar =1$, $\tilde{H}$ reduces
to $\hat{H}$ for $\alpha _{1}\rightarrow A$, $\alpha _{2}\rightarrow -iA$, $%
\alpha _{3}\rightarrow iB$ and $\alpha _{4}\rightarrow B$. We also notice
that $\mathcal{PT}_{-}$ is no longer a symmetry of $\hat{H}\,$, i.e. $[%
\mathcal{PT}_{-},\hat{H}]\neq 0$.

Our aim here is to employ ${\mathcal{D}}$-pbs to diagonalize $\tilde{H}$
exactly, instead of using a perturbative approach as in \cite{margh,miao2}
and to determine its spectrum. For this purpose we convert the Hamiltonian
first from a flat noncommutative space to one in terms of standard canonical
variables $x_{i}$ and $p_{i}$ for $i=1,2$ satisfying the canonical
commutation relations $\left[ x_{j},p_{j}\right] =i\hbar $ and $\left[
x_{i},x_{j}\right] =\left[ p_{i},p_{j}\right] =0$. This is achieved by a
standard Bopp shift $\tilde{x}_{1}\rightarrow x_{1}-\frac{\theta }{2\hbar }%
p_{2}$, $\tilde{x}_{2}\rightarrow x_{2}+\frac{\theta }{2\hbar }p_{1}$,\ $%
p_{1}\rightarrow p_{1}$ and $p_{2}\rightarrow p_{2}$. The Hamiltonian in (%
\ref{HL}) then acquires the form%
\begin{eqnarray}
\tilde{H} &=&\left( \frac{1}{2m}+\frac{m\omega ^{2}\theta ^{2}}{8\hbar ^{2}}%
\right) (p_{1}^{2}+p_{2}^{2})+\frac{m\omega ^{2}}{2}(x_{1}^{2}+x_{2}^{2})+%
\frac{m\omega ^{2}\theta }{2\hbar }\left( x_{2}p_{1}-x_{1}p_{2}\right) ~~
\label{H1} \\
&&+i\alpha _{1}x_{1}+\alpha _{2}x_{2}+\left( \alpha _{3}+\frac{\alpha
_{2}\theta }{2\hbar }\right) p_{1}+i\left( \alpha _{4}-\frac{\alpha
_{1}\theta }{2\hbar }\right) p_{2}.  \notag
\end{eqnarray}%
We now attempt to re-express this Hamiltonian in terms of pseudo-bosonic
number operators $N_{i}=b_{i}a_{i}$ as%
\begin{equation}
\tilde{H}=\gamma _{1}N_{1}+\gamma _{2}N_{2}+\gamma _{0}\qquad \text{for }%
\gamma _{0},\gamma _{1},\gamma _{2}\in \mathbb{R},  \label{ps}
\end{equation}%
where the operators $a_{i}$ and $b_{i}$ obey the two dimensional
pseudo-bosonic commutation relations%
\begin{equation}
\left[ a_{j},b_{k}\right] =i\delta _{jk},\qquad \left[ a_{j},a_{k}\right] =%
\left[ b_{j},b_{k}\right] =0,\qquad \text{for }j,k=1,2.
\end{equation}%
For this purpose we represent the pseudo-bosonic operators $a_{i}$ and $%
b_{i} $ in terms of standard bosonic creation and annihilation operators $%
A_{i}^{\dagger }$ and $A_{i}$, respectively, 
\begin{eqnarray}
a_{1} &=&\frac{1}{\sqrt{2}}(A_{1}+iA_{2})+i\beta _{1},\qquad b_{1}=\frac{1}{%
\sqrt{2}}(A_{1}^{\dagger }-iA_{2}^{\dagger })+i\beta _{3},  \label{pb1} \\
a_{2} &=&-\frac{1}{\sqrt{2}}(iA_{1}+A_{2})+\beta _{2},\qquad b_{2}=\frac{1}{%
\sqrt{2}}(iA_{1}^{\dagger }-A_{2}^{\dagger })+\beta _{4},  \label{pb2}
\end{eqnarray}%
with $[A_{j},A_{k}^{\dagger }]=i\delta _{jk}$, $\left[ A_{j},A_{k}\right]
=[A_{j}^{\dagger },A_{k}^{\dagger }]=0$ for $j,k=1,2$ and $\beta _{i}\in 
\mathbb{C}$ for $i=1,2,3,4$. Furthermore we represent the $A_{i}^{\dagger }$
and $A_{i}$ in terms of the standard canonical variables%
\begin{eqnarray}
A_{1} &=&\sqrt{\frac{M\omega }{2\hbar }}x_{1}+i\sqrt{\frac{1}{2\hbar M\omega 
}}p_{1},\qquad A_{2}=\sqrt{\frac{M\omega }{2\hbar }}x_{2}+i\sqrt{\frac{1}{%
2\hbar M\omega }}p_{2}, \\
A_{1}^{\dagger } &=&\sqrt{\frac{M\omega }{2\hbar }}x_{1}-i\sqrt{\frac{1}{%
2\hbar M\omega }}p_{1},\qquad A_{2}^{\dagger }=\sqrt{\frac{M\omega }{2\hbar }%
}x_{2}-i\sqrt{\frac{1}{2\hbar M\omega }}p_{2}.
\end{eqnarray}%
We note that the pseudo-bosonic operators reduce to standard boson operators
with $b_{i}=a_{i}^{\dagger }$ if and only if for $\beta _{1}=-\bar{\beta}%
_{3} $ and $\beta _{2}=\bar{\beta}_{4}$. Upon substitution we compare now (%
\ref{ps}) and (\ref{H1}), which become identical subject to the constraints%
\begin{eqnarray}
\beta _{1} &=&\frac{\Omega (\alpha _{1}+\alpha _{2})+2\hbar m\omega (\alpha
_{3}-\alpha _{4})}{\left( \Omega +\theta m\omega \right) \sqrt{2m\Omega
\omega ^{3}}},\qquad \beta _{2}=\frac{\Omega (\alpha _{1}-\alpha
_{2})+2\hbar m\omega (\alpha _{3}+\alpha _{4})}{\left( \Omega -\theta
m\omega \right) \sqrt{2m\Omega \omega ^{3}}},  \label{c1} \\
\beta _{3} &=&\frac{\Omega (\alpha _{1}-\alpha _{2})-2\hbar m\omega (\alpha
_{3}+\alpha _{4})}{\left( \Omega +\theta m\omega \right) \sqrt{2m\Omega
\omega ^{3}}},\qquad \beta _{4}=\frac{-\Omega (\alpha _{1}+\alpha
_{2})+2\hbar m\omega (\alpha _{3}-\alpha _{4})}{\left( \Omega -\theta
m\omega \right) \sqrt{2m\Omega \omega ^{3}}},~~~~  \label{c2} \\
\gamma _{0} &=&\frac{1}{2}\omega \left[ \Omega \left( 1+\beta _{1}\beta
_{3}-\beta _{2}\beta _{4}\right) +\theta m\omega \left( \beta _{1}\beta
_{3}+\beta _{2}\beta _{4}\right) \right] , \\
\gamma _{1} &=&\frac{1}{2}\omega \left( \Omega +\theta m\omega \right)
,\qquad \gamma _{2}=\frac{1}{2}\omega \left( \Omega -\theta m\omega \right)
,\qquad M=\frac{2m\hbar }{\Omega },  \label{c4}
\end{eqnarray}%
where $\Omega :=\sqrt{4\hbar ^{2}+\theta ^{2}m^{2}\omega ^{2}}$. If we are
now able to construct eigenstates $\Psi _{\underline{n}}$ for the
pseudo-bosonic number operators such that $N_{i}\varphi _{\underline{n}%
}=\hbar \omega n_{i}\varphi _{\underline{n}}$, the eigenvalues for $\tilde{H}
$ are immediately computed from (\ref{ps}) to 
\begin{equation}
E_{n_{1},n_{2}}=\gamma _{1}\hbar \omega n_{1}+\gamma _{2}\hbar \omega
n_{2}+\gamma _{0}.  \label{E}
\end{equation}

We observe from (\ref{c1}) to (\ref{c4}) that the constants $\gamma _{i}\in 
\mathbb{R}$ for $i=0,1,2$ are real and consequently the energy $%
E_{n_{1},n_{2}}$ is also real. Furthermore, we observe that the presence of
the linear terms in (\ref{HL}), that is $\alpha _{i}\neq 0$ for $i=1,2,3,4$,
prevents us from using a standard bosonic oscillator algebra and we are
forced to employ pseudo-bosons. This is seen from the fact that the
pseudo-bosonic operator reduce to standard boson operators if and only if
for $\beta _{1}=-\bar{\beta}_{3}$ and $\beta _{2}=\bar{\beta}_{4}$. However,
our constraints (\ref{c1}) and (\ref{c2}) imply that in this boson case some
linear terms in our Hamiltonian have to vanish, that is $\alpha _{1}=\alpha
_{4}=0$.

Furthermore we notice that for the reduction of $\tilde{H}$ to $\hat{H}$ for 
$\alpha _{1}\rightarrow A$, $\alpha _{2}\rightarrow iA$, $\alpha
_{3}\rightarrow iB,\alpha _{4}\rightarrow B$ we obtain $\beta _{1}=\bar{\beta%
}_{3}$ and $\beta _{2}=-\bar{\beta}_{4}$, such that $\gamma _{0}$ and
therefore $E_{n_{1},n_{2}}$ remain real. In this case the $\mathcal{PT}_{-}$%
-symmetry is broken and it remains unclear which antilinear symmetry, if
any, is responsible for keeping the spectrum real.

Let us now verify that eigenstates $\varphi _{\underline{n}}$ and those of
the adjoint of the Hamiltonian, $\Psi _{\underline{n}}$, are well defined,
really exist and most crucially whether they constitute a Riesz basis, or
even a basis.

\subsection{Verification of the pseudo-bosonic assumptions}

For simplicity let us now adopt atomic units. We commence by introducing the
operators%
\begin{equation}
\hat{a}_{i}:=\lim\nolimits_{\beta _{i}\rightarrow 0}a_{i},\qquad \hat{a}%
_{i}^{\dagger }:=\lim\nolimits_{\beta _{i}\rightarrow 0}b_{i},
\end{equation}%
which, from (\ref{pb1})-(\ref{pb2}), satisfy the standard bosonic canonical
commutation relations, $[\hat{a}_{i},\hat{a}_{j}^{\dagger }]=\delta
_{i,j}\,1\!\!1$, $[\hat{a}_{i},\hat{a}_{j}]=0$, for $i,j=1,2$. Then,
introducing the unitary operators 
\begin{equation}
D_{i}(z):=\exp \left\{ \overline{z}\,\hat{a}_{i}-z\,\hat{a}_{i}^{\dagger
}\right\} ,\qquad D(\underline{z}):=D_{1}(z_{1})D_{2}(z_{2}),
\end{equation}%
we compute%
\begin{equation}
a_{i}=\hat{a}_{i}+\nu _{i}=D(\underline{\nu })\hat{a}_{i}D^{-1}(\underline{%
\nu }),\qquad b_{i}=\hat{a}_{i}^{\dagger }+\mu _{i}=D(\underline{\mu })\hat{a%
}_{i}^{\dagger }D^{-1}(\underline{\mu }),  \label{DD}
\end{equation}%
for $i=1,2$ with $\underline{\nu }:=\{i\beta _{1},\beta _{2}\}$, $\underline{%
\mu }:=\{-i\bar{\beta}_{3},\bar{\beta}_{4}\}$. An orthonormal basis for $%
\mathcal{H}=\mathcal{L}^{2}(\mathbb{R}^{2})$ is then constructed easily: Let 
$e_{0,0}=e_{\underline{0}}$ be the vacuum of $\hat{a}_{1}$ and $\hat{a}_{2}$%
, that is $\hat{a}_{i}e_{\underline{0}}=0$ for $i=1,2$. Then as common for
the purely bosonic case, we introduce 
\begin{equation}
e_{n_{1},\,n_{2}}=e_{\underline{n}}:=\frac{1}{\sqrt{n_{1}!n_{2}!}}(\hat{a}%
_{1}^{\dagger })^{n_{1}}(\hat{a}_{2}^{\dagger })^{n_{2}}e_{\underline{0}},
\end{equation}%
and the related orthonormal basis $\mathcal{F}_{e}=\{e_{\underline{n}%
},\,n_{1},n_{2}\geq 0\}$. Of course for the bosonic number operator $\hat{n}%
_{i}:=\hat{a}_{i}^{\dagger }\hat{a}_{i}$ we have $\hat{n}_{i}e_{\underline{n}%
}=n_{i}e_{\underline{n}}$.

In order to verify the assumptions of section 2, we first seek to construct $%
\varphi _{\underline{0}}$, i.e. the vacuum of $a_{i}$ satisfying $%
a_{1}\varphi _{\underline{0}}=a_{2}\varphi _{\underline{0}}=0$. Evidently
this holds if, and only if, $\hat{a}_{i}(D^{-1}(\underline{\nu })\varphi _{%
\underline{0}})=0$ for $i=1,2$. This implies that $\varphi _{\underline{0}%
}=D(\underline{\nu })e_{\underline{0}}$, up to a normalization which will be
fixed below. Notice that, due to fact that $D(\underline{\nu })$ is unitary,
and therefore everywhere defined, $\varphi _{\underline{0}}$ is well defined.

Similarly we derive $\Psi _{\underline{0}}$, the vacuum for $b_{j}^{\dagger
} $. We require $b_{1}^{\dagger }\Psi _{\underline{0}}=b_{2}^{\dagger }\Psi
_{\underline{0}}=0$ which can be rewritten as $\hat{a}_{i}(D^{-1}(\underline{%
\mu })\Psi _{\underline{0}})=0$ for $i=1,2$. These equations are solved by $%
\Psi _{\underline{0}}=N_{\Psi }D(\underline{\mu })e_{\underline{0}}$, which,
due to the unitarity of $D(\underline{\mu }) $ is again well defined. Here $%
N_{\Psi }$ is a normalization needed to ensure the normalization $%
\left\langle \varphi _{\underline{0}},\Psi _{\underline{0}}\right\rangle =1$%
. It is computed to 
\begin{equation}
N_{\Psi }^{2}=\frac{\left\langle \varphi _{\underline{0}},\varphi _{%
\underline{0}}\right\rangle }{\left\langle \Psi _{\underline{0}},\Psi _{%
\underline{0}}\right\rangle }=\exp \left[ \left\vert \beta _{1}\right\vert
^{2}+\left\vert \beta _{2}\right\vert ^{2}-\left\vert \beta _{3}\right\vert
^{2}-\left\vert \beta _{4}\right\vert ^{2}-2\func{Re}(\beta _{1}\beta _{2})-2%
\func{Re}(\beta _{3}\beta _{4})\right] .
\end{equation}%
Evidently for $\beta _{2}=-\beta _{3}$ and $\beta _{1}=\beta _{4}$ this
reduces to the standard bosonic normalization, as is expected.\vspace{2mm}

\textbf{Remark: }These results could have also been found quite easily by
solving the equations directly in the coordinate representation. For
instance, $a_{1}\varphi _{\underline{0}}=a_{2}\varphi _{\underline{0}}=0$
are equivalent to the differential equations 
\begin{equation}
\left( x_{1}+\partial _{x_{1}}+ix_{2}+i\,\partial _{x_{2}}+2i\beta
_{1}\right) \varphi _{\underline{0}}(x_{1},x_{2})=\left( -ix_{1}-i\,\partial
_{x_{1}}-x_{2}-\partial _{x_{2}}+2\beta _{2}\right) \varphi _{\underline{0}%
}(x_{1},x_{2})=0,
\end{equation}%
solved by $\varphi _{\underline{0}}(x_{1},x_{2})\propto e^{-\frac{1}{2}%
(x_{1}^{2}+x_{2}^{2})-i(\beta _{1}+\beta _{2})x_{1}-(\beta _{1}-\beta
_{2})x_{2}}$. Similarly we find $\Psi _{\underline{0}}(x_{1},x_{2})\propto
e^{-\frac{1}{2}(x_{1}^{2}+x_{2}^{2})+i(\beta _{3}-\beta _{4})x_{1}+(\beta
_{3}+\beta _{4})x_{2}}$. We see that both of these functions belong, for
instance, to the set $\mathcal{S}(\mathbb{R}^{2}\mathbb{)}$ of $C^{\infty }$%
-functions which, together with their derivatives, decrease faster to zero
than any inverse power of $x_{1}$ and $x_{2}$. However, this property might
not be enough for our purposes, since as we have outlined in section 2, we
need to identify a set ${\mathcal{D}}$, dense in $\mathcal{H}$, which not
only contains $\varphi _{\underline{0}}$ and $\Psi _{\underline{0}}$, but
which is in addition also stable under the action of $a_{j}^{\sharp }$, $%
b_{j}^{\sharp }$, and other relevant operators. It is convenient to
introduce, therefore, the following set: 
\begin{equation}
{\mathcal{D}}=\left\{ f(x_{1},x_{2})\in \mathcal{S}(\mathbb{R}^{2}),\,%
\mbox{ such
that }e^{k_{1}x_{1}+k_{2}x_{2}}f(x_{1},x_{2})\in \mathcal{S}(\mathbb{R}%
^{2}),\,\forall k_{1},k_{2}\in \mathbb{C}\right\} .
\end{equation}%
${\mathcal{D}}$ is dense in $\mathcal{H}$, since it contains the set $D(%
\mathbb{R}^{2}\mathbb{)}$ of the $C^\infty$-functions with compact support.

\vspace{2mm}

Following section 2, we are now interested in deducing the properties of the
vectors $\varphi _{\underline{n}}=\frac{1}{\sqrt{n_{1}!n_{2}!}}%
b_{1}^{n_{1}}b_{2}^{n_{2}}\varphi _{\underline{0}}$ and $\Psi _{\underline{n}%
}=\frac{1}{\sqrt{n_{1}!n_{2}!}}(a_{1}^{\dagger })^{n_{1}}(a_{2}^{\dagger
})^{n_{2}}\Psi _{\underline{0}}$. We notice that both $\varphi _{\underline{n%
}}$ and $\Psi _{\underline{n}}$ necessarily belong to ${\mathcal{D}}$ for
all $\underline{n}$, because of the stability of ${\mathcal{D}}$ under the
action of $b_{i}$ and $a_{i}^{\dagger }$, and the previously established
fact that $\varphi _{\underline{0}},\Psi _{\underline{0}}\in {\mathcal{D}}$.
The formulae (\ref{DD}) state how the pseudo-bosonic operators $%
(a_{i},b_{i}) $ are related to the bosonic operators $(\hat{a}_{i},\,\hat{a}%
_{i}^{\dagger })$ by means of the in general two different unitary operators 
$D(\underline{\nu })$ and $D(\underline{\mu })$.

\bigskip A single operator could be used if we introduce the operators 
\begin{equation}
V_{i}(z,w):=\exp \left\{ \bar{w}\,\hat{a}_{i}-z\,\hat{a}_{i}^{\dagger
}\right\} ,\qquad V(\underline{\nu },\underline{\mu }):=V_{1}(\nu _{1},\mu
_{1})V_{2}(\nu _{2},\mu _{2}).
\end{equation}
Now we compute 
\begin{equation}
a_{i}=V(\underline{\nu },\underline{\mu })\hat{a}_{i}V^{-1}(\underline{\nu },%
\underline{\mu }),\qquad b_{i}=V(\underline{\nu },\underline{\mu })\hat{a}%
_{i}^{\dagger }V^{-1}(\underline{\nu },\underline{\mu }),  \label{412}
\end{equation}%
which, in contrast to (\ref{DD}), only involve a single, albeit in general
unbounded, operator to relate the $(a_{i},b_{i})$ to the $(\hat{a}_{i},\,%
\hat{a}_{i}^{\dagger })$. We also check directly 
\begin{equation}
a_{i}^{\dagger }=V(\underline{\mu },\underline{\nu })\hat{a}_{i}V^{-1}(%
\underline{\mu },\underline{\nu }),\qquad b_{i}^{\dagger }=V(\underline{\mu }%
,\underline{\nu })\hat{a}_{i}V^{-1}(\underline{\mu },\underline{\nu }).
\label{413}
\end{equation}%
A immediate consequence of these formulae are the following relations
between the various number operators: $\hat{n}_{i}=V^{-1}(\underline{\nu },%
\underline{\mu })N_{i}V(\underline{\nu },\underline{\mu })=V^{-1}(\underline{%
\mu },\underline{\nu })N_{i}^{\dagger }V(\underline{\mu },\underline{\nu })$%
, which in turns implies that 
\begin{equation}
N_{i}=T(\underline{\nu },\underline{\mu })N_{i}^{\dagger }T^{-1}(\underline{%
\nu },\underline{\mu }),  \label{414}
\end{equation}%
where $T(\underline{\nu },\underline{\mu }):=V(\underline{\nu },\underline{%
\mu })V^{-1}(\underline{\mu },\underline{\nu })$. Needless to say, all these
equalities and definitions are well defined on ${\mathcal{D}}$, but not on
the whole $\mathcal{H}$\footnote{%
This aspect is almost never stressed in the physical literature. Unbounded
operators never exist \emph{alone}! They exist in connection with some
suitable dense subspace of $\mathcal{H}$, their domains.}. Incidentally we
observe that $T(\underline{\gamma },\underline{\gamma })=1\!\!1$. This is in
agreement with the fact that, when $\underline{\mu }=\underline{\nu }$, the
operator $V(\underline{\nu },\underline{\mu })$ is bounded with bounded
inverse, see below.

By a similar reasoning as above applied for the construction of the vacuum
state we now deduce that 
\begin{equation}
\varphi _{\underline{n}}=V(\underline{\nu },\underline{\mu })e_{\underline{n}%
},\qquad \Psi _{\underline{n}}=N_{\Psi }V(\underline{\mu },\underline{\nu }%
)e_{\underline{n}}.  \label{411}
\end{equation}%
In analogy with \cite{bag2013august}, we see that, while $V(\underline{\nu },%
\underline{\nu })=D(\underline{\nu })$ is a unitary operator and as a
consequence bounded, the operator $V(\underline{\nu },\underline{\mu })$, as
well as its inverse, is unbounded for $\underline{\nu }\neq \underline{\mu }$%
. The crucial conclusion from this is that the two sets $\mathcal{F}%
_{\varphi }=\{\varphi _{\underline{n}}\}$ and $\mathcal{F}_{\Psi }=\{\Psi _{%
\underline{n}}\}$ cannot be Riesz bases. In fact, they are both related to
the orthonormal basis $\mathcal{F}_{e}$ by unbounded operators. Moreover:
they are not even a basis, while they are both complete in $\mathcal{H}$.
The proofs of these claims do not differ much from those given in \cite%
{bag2013august} and therefore will not be repeated here. We will comment
further on the physical meaning of these results in the next subsection.

Similarly as in \cite{bag2013august} we can prove that $\mathcal{F}_{\varphi
}$ and $\mathcal{F}_{\Psi }$ are ${\mathcal{D}}$-quasi bases. In fact,
repeating almost the same steps, we deduce that for instance, $\forall
\,f,g\in {\mathcal{D}}$, 
\begin{equation}
\left\langle f,g\right\rangle =\sum\nolimits_{\underline{n}}\left\langle
f,\varphi _{\underline{n}}\right\rangle \left\langle \Psi _{\underline{n}%
},g\right\rangle,
\end{equation}%
so that the results listed at the end of section 2 hold true. In particular,
let us introduce the operator $\Theta (\underline{\nu },\underline{\mu }):=T(%
\underline{\mu },\underline{\nu })$. It is possible to show that $\Theta (%
\underline{\nu },\underline{\mu })$ is self-adjoint, invertible, and leaves $%
{\mathcal{D}}$ invariant. Moreover, $\Theta (\underline{\nu },\underline{\nu 
})=1$, and 
\begin{equation}
\Theta (\underline{\nu },\underline{\mu })=N_{\Psi }\prod_{i=1}^{2}e^{({\nu }%
_{i}-{\mu }_{i})\hat{a}_{i}^{\dagger }}e^{({\bar{\nu}}_{i}-{\bar{\mu}}_{i})%
\hat{a}_{i}},
\end{equation}%
which implies that $\left\langle f,\Theta (\underline{\nu },\underline{\mu }%
)f\right\rangle >0$ for all non zero vectors $f\in {\mathcal{D}}$. This is
in agreement with the facts that (i) $\Psi _{\underline{n}}=\Theta (%
\underline{\nu },\underline{\mu })\varphi _{\underline{n}}$, $\forall \,%
\underline{n}$; (ii) $(a_{j},b_{j}^{\dagger })$ are $\Theta $-conjugate: $%
a_{j}f=\Theta ^{-1}(\underline{\nu },\underline{\mu })b_{j}^{\dagger }\Theta
(\underline{\nu },\underline{\mu })f$, for all $f\in {\mathcal{D}}$. We
conclude also that, again for all $f\in {\mathcal{D}}$, 
\begin{equation}
N_{i}f=\Theta ^{-1}(\underline{\nu },\underline{\mu })N_{i}^{\dagger }\Theta
(\underline{\nu },\underline{\mu })f,
\end{equation}
which is the intertwining relation responsible for the fact that $\tilde H$
and $\tilde H^\dagger$ have the same eingenvalues and related eigenvectors,
see below.

\subsection{Back to the Hamiltonian}

Let us now return to our original problem, i.e. the deduction of the
eigenvalues and the eigenvectors for $\tilde{H}$ in (\ref{H1}) and $\hat{H}$
in (\ref{1}). As we have shown we may express them in terms of
pseudo-bosonic number operators. From the above construction is clear that 
\begin{equation}
\tilde{H}\varphi _{\underline{n}}=E_{\underline{n}}\varphi _{\underline{n}},
\label{415}
\end{equation}%
with $E_{\underline{n}}\in \mathbb{R}$ given by (\ref{E}). From our results
in section 2 it also follows directly that the eigensystem of the adjoint $%
\tilde{H}^{\dagger }=\bar{\gamma}_{1}N_{1}^{\dagger }+\bar{\gamma}%
_{2}N_{2}^{\dagger }+\bar{\gamma}_{0}$ is computed to 
\begin{equation}
\tilde{H}^{\dagger }\Psi _{\underline{n}}=\bar{E}_{\underline{n}}\Psi _{%
\underline{n}}=E_{\underline{n}}\Psi _{\underline{n}}.  \label{416}
\end{equation}

The analysis in \cite{bag2013august} showed that, as already deduced, two
biorthogonal sets of eigenstates of a Hamiltonian and of its adjoint, need
not to be automatically a Riesz basis, even when they are complete! This is
exactly the case here: $\mathcal{F}_{\varphi }$ and $\mathcal{F}_{\Psi }$
are biorthogonal, complete, eigenstates of $\tilde{H}$ and $\tilde{H}%
^{\dagger }$ ($\hat{H}$ and $\hat{H}^{\dagger }$), respectively, but neither 
$\mathcal{F}_{\varphi }$ nor $\mathcal{F}_{\Psi }$ are bases for $\mathcal{H}
$. However, interestingly enough, they are ${\mathcal{D}}$-quasi bases, and
this is reflected in the properties we have explicitly verified for our
model.

\section{Conclusions}

We have investigated the properties of a non self-adjoint model on a
noncommutative two dimensional space. The Hamiltonian $\tilde{H}$ was set up
in the standard fashion followed in the literature on $\mathcal{PT}$%
-symmetric quantum mechanics, by seeking an anti-linear symmetry, i.e. $%
\mathcal{PT}_{-}$ in this case. From our explicit formulae we observe that $%
\mathcal{PT}_{-}$: $\varphi _{\underline{0}}\rightarrow \varphi _{\underline{%
0}}$, $\varphi _{\underline{n}}\rightarrow (-1)^{n_{1}}\varphi _{\underline{n%
}}$, $\Psi _{\underline{0}}\rightarrow \Psi _{\underline{0}}$, $\Psi _{%
\underline{n}}\rightarrow (-1)^{n_{1}}\Psi _{\underline{n}}$ such that by
the standard arguments of Wigner \cite{EW} it follows that the eigenvalues
of $\tilde{H}$ have to be real. This is confirmed by our explicit
computation. The symmetry for the Hamiltonian $\hat{H}$ is not evident from
the start, but as demonstrated the overall conclusions are the same as for $%
\tilde{H}.$

However, despite having well defined real physical spectrum, we established
further that $\tilde{H}$ can not be considered as a standard quantum
mechanical model, since the corresponding biorthonormal system is not of
Riesz type. As already discussed, in many places in the literature, see \cite%
{probl1} for instance, it is incorrectly assumed that the eigenvectors of a
not self-adjoint Hamiltonian $H$ and $H^{\dagger }$ automatically form a
biorthogonal basis. In fact, this is a rather strong requirement which is
quite difficult to find satisfied in concrete models existing in the
literature, at least for infinite dimensional Hilbert spaces. We have shown
that even for the simple example presented here this is not the case. This
only leaves two of the following options: either this conclusion is wrong
for the cases treated, as it would be for the model presented here, or at
least some additional analysis is required to justify it. Thus our example
supports the suggestion \cite{petr,bag2013august} that many models, thought
to be very interesting quantum mechanical systems, need to be revisited for
further scrutiny.

It is easy to see from our formulae that these conclusions do not rely on
the fact that the model is formulated on a noncommutative space and also
hold in the limit to the commutative space when setting $\lim_{\theta
\rightarrow 0}\Omega =2\hbar $, $\lim_{\theta \rightarrow 0}M=m$, etc. In
reverse, this also means that the problem of not having automatically a
biorthonormal basis can not be solved by formulating the model on a
non-commutative space, which provides more freedom and often removes
inconsistencies.

We end this section, and the paper, observing that, even with all the
problems we have put in evidence along the paper, we may still make sense of
the model presented here, simply because of the role of the quasi-bases as
described above and in more detail in the quoted literature.

\section*{Acknowledgements}

This work was partially supported by the University of Palermo and in part
from INFN, Torino.

\end{document}